\documentclass[journal]{IEEEtran}

\usepackage{xr}
\usepackage{moreverb,url}
\usepackage[colorlinks,bookmarksopen,bookmarksnumbered,citecolor=black,urlcolor=black]{hyperref}
\usepackage{amsmath}
\usepackage{graphicx}
\usepackage{xcolor}
\usepackage{subfig}
\usepackage{booktabs}
\begin{document}

\title{Comparison of the LASSO and Integrative LASSO with Penalty Factors (IPF-LASSO) methods for multi-omics data: Variable selection with Type I error control}

\author{
Charlotte Castel,
Zhi Zhao,
 Magne Thoresen
\thanks{C. Castel, Postdoctoral Researcher, Z. Zhao, Researcher, and M. Thoresen, Professor, are with the Centre for Biostatistics and Epidemiology (OCBE), Institute of Basic Medical Sciences, Faculty of
Medicine, University of Oslo, Oslo, Norway (e-mail: c.j.y.castel@medisin.uio.no; zhi.zhao@medisin.uio.no; magne.thoresen@medisin.uio.no).} 
}

\maketitle

\begin{abstract}
Variable selection in relation to regression modeling has constituted a methodological problem for more than 60 years. Especially in the context of high-dimensional regression, developing stable and reliable methods, algorithms, and computational tools for variable selection has become an important research topic. Omics data is one source of such high-dimensional data, characterized by diverse genomic layers, and an additional analytical challenge is how to integrate these layers into various types of analyses. While the IPF-LASSO model has previously explored the integration of multiple omics modalities for feature selection and prediction by introducing distinct penalty parameters for each modality, the challenge of incorporating heterogeneous data layers into variable selection with Type I error control remains an open problem. To address this problem, we applied stability selection as a method for variable selection with false positives control in both IPF-LASSO and regular LASSO. The objective of this study was to compare the LASSO algorithm with IPF-LASSO, investigating whether introducing different penalty parameters per omics modality could improve statistical power while controlling false positives. Two high-dimensional data structures were investigated, one with independent data and the other with correlated data. The different models were also illustrated using data from a study on breast cancer treatment, where the IPF-LASSO model was able to select some highly relevant clinical variables.

\end{abstract}


\section{Introduction}
Variable selection in regression models has been an important methodological challenge for more than 60 years \cite{George2000}. Developing stable and dependable methods, algorithms, and computational tools for variable selection stands as a crucial task, especially in the challenging context of high-dimensional regression, where the number of potential explanatory variables exceeds the number of observations (p $\gg$ n). The conventional approach assumes sparsity, implying that only a small subset of the explanatory variables genuinely influences the outcome. Methods like the LASSO\cite{tibshirani1996regression} have become very popular for integrating variable selection into the estimation processes. 

A hot topic in omics research during recent years 
 is integration of different layers of genomic information. Due to technological developments, we often have available diverse high-dimensional measurements such as DNA methylation, copy number variation, and mRNA expression for the same individual, in addition to more traditional clinical data. A structured integration of such different layers gives an opportunity to gain a holistic understanding of the biological system in question.
 For prediction purposes, the IPF-LASSO \cite{boulesteix2017ipf}, the Priority-Lasso \cite{klau2018priority}, the IPFStructPenalty \cite{zhao2020structured}, the mix-lasso \cite{zhao2022tissue} and the cooperative learning\cite{ding2022cooperativ} methods have been suggested. They all perform variable selection, but none of them incorporate any type of error control. Thus, the challenge of integrating heterogeneous data layers into methods for variable selection with Type I error control (false positive) remains an open problem. 
 
The IPF-LASSO model was initially developed in 2017 to address the integration of multiple omics modalities for variable selection in predictive modeling. Boulesteix et al. \cite{boulesteix2017ipf} proposed a penalized regression method to address this challenge, applying distinct penalty factors for different data modalities.
The results in this article indicate that the IPF-LASSO method and regular LASSO are comparable in terms of prediction accuracy, but with fewer variables selected for the IPF-LASSO. These results would indicate that the IPF-LASSO could select variables with fewer false positives compared to the regular LASSO, which again begs the question if the IPF-LASSO would achieve higher power under proper error control, compared to the regular LASSO. 

Managing error control in statistical analyses can be done in various ways; by controlling the false discovery rate (FDR), controlling the family-wise error rate (FWR), or controlling the number of false positives. To our knowledge, there are four main approaches for applying error control in variable selection in high-dimensional regression models.(i) Stability selection \cite{meinshausen2010stability,shah2013variable} aggregates results from applying a selection procedure to subsamples, controlling the expected number of false positives. (ii) Data splitting methods \cite{cox1975note,wasserman2009high,meinshausen2009p,dai2022false,rasines2021splitting}, controlling FDR or FWR, dividing the data randomly into two parts for variable selection and inference, respectively. (iii) Selective inference \cite{fithian2014optimal,lockhart2014significance,lee2016exact} also controls FDR or FWR, by formally conditioning on the selection event. (iv) Model-X knockoff and its like \cite{candes2018panning} generates knockoff predictors to mimic the covariance pattern of the original predictors, controlling FDR in variable selection. The work presented here focuses on the control of the number of false positives. 
\newline
\indent To our knowledge, there is no existing work on variable selection with error control in high-dimensional models integrating multiple omics modalities. Stability selection is a widely used method for variable selection with control over the expected number of false positives. We applied stability selection to the IPF-LASSO. The objective was to compare the popular LASSO algorithm with the IPF-LASSO to evaluate if introducing different penalty factors per data modality improves power while controlling the number of false positives, compared to the standard LASSO. Two high-dimensional data settings were investigated: one setting with independent variables and one setting with correlated variables. The simulation design is the same as in the IPF-LASSO article \cite{boulesteix2017ipf}, determining the best penalty parameters through cross-validation (CV). 
The focus in this article was on binary outcomes and logistic models, and we will keep the same focus, although the methods are of course equally relevant also for other models.
\newline
\indent In our simulation studies, we observe that both the IPF-LASSO and the LASSO control the number of false positives well; however, the IPF-LASSO increases the power compared to standard LASSO, particularly in scenarios with a small true model size, a high absolute difference between proportions of relevant variables in the two modalities, and a small ratio between the size of the smaller modality and the larger modality.
For a real data application, we chose to test our models for variable selection with error control on the same breast cancer dataset from the study by Hatzis et al. \cite{hatzis2011genomic}, as utilized in the article by Boulesteix et al. \cite{boulesteix2017ipf} This dataset includes a relatively small set of clinical variables in addition to a high-dimensional set of gene expressions, which constitutes our two data  modalities. 
\newline
\indent This article is divided as follows. After this introduction, Section 2 describes the different methods, Section 3 presents the results from the simulation studies, and Section 4 presents the results from the analysis of the breast cancer data. In Section 5 we discuss our findings and conclude.

\section{Methods}

\subsection{Integrative LASSO with Penalty Factors (IPF-LASSO)}
Let $x_{ij}$ represent the predictor variable $j$ measured on subject $i$, and $y_{i}$ represent the (continuous) response values associated, where $i = 1, . . . ,n$ and $j= 1, . . . , p$.
The idea of the standard LASSO method for the linear model involves solving the $\ell_1$-penalized regression problem by determining $\boldsymbol{\beta}$ = $\left\{\beta_j \right\}$ that minimizes the loss function:
\begin{center}
    \[ \sum_{i=1}^{n} \left(y_i - \sum_{j=1}^p x_{ij} \beta_j \right)^2 + \lambda \lVert \boldsymbol{\beta} \rVert_1  ,\]
\end{center}
where the $\ell_1$-norm is denoted by  \[ \lVert . \rVert _{1}= \sum_{j=1}^p |\beta_{j}| \]
and $\lambda$ is the penalty parameter.\\
\newline
Readers can explore Tibshirani's work for more information about LASSO regression \cite{tibshirani1996regression}. The LASSO framework has been extended to other types of regression models, e.g. logistic regression \cite{friedman2010regularization} which we will take advantage of in this work\\

Boulesteix et al. \cite{boulesteix2017ipf} proposed to deal with the case of multiple data modalities by introducing a small modification to the  classical LASSO. 
Let us denote by $X_1^{(m)}$, ... , $X_{p_m}^{(m)}$ the variables belonging to modality $m$ (for $m = 1$,...,$M$), with their respective observed values $x_{i1}^{(m)}$, ..., $x_{ip_m}^{(m)}$ for subject $i$ (for $i = 1, . . . ,n$). The value $p_m$ corresponds to the number of variables within modality $m$. In the same way, $\beta_j^{(m)}$ $(j=1,...,p_m)$ represents the regression parameters associated with variable $X_{j}^{(m)}$  from modality $m$. 
The idea of the IPF-LASSO is to use a weighted sum of the $l_{1}$-norm of the coefficient vector of each data modality as the penalty term to take into account the differences between data modalities. 
The new $l_{1}$-penalized regression problem is solved by minimizing
\begin{center}
    \[ \sum_{i=1}^{n} \left(y_i - \sum_{m=1}^M \sum_{j=1}^{p_m} x_{ij}^{(m)} \beta_j^{(m)} \right)^2 + \sum_{m=1}^M \lambda_m \lVert \boldsymbol{\beta}^{(m)} \rVert_1  \]
\end{center}
with $\lambda_m > 0$ being the penalty parameter applied to the variables from modality $m$. How to choose $\lambda_m$ for $m = 1, . . . ,M$ becomes a challenging task. Boulesteix et al. introduce the penalty factors to solve this problem. 
As explained in Boulesteix et al., the term ``penalty factors" refers to the multiplicative factors applied to the penalty term. For simplicity, the reference modality is the first modality with penalty $\lambda_1$ and associated penalty factor 1. The penalty factor for any given modality $\lambda_m$ is then determined by the ratio $\lambda_m/\lambda_1$.
The estimation, the connection with the other LASSO variations and the choice of the penalty parameters are well explained in the work of Boulesteix et al.\cite{boulesteix2017ipf}.\\

\subsection{Stability selection}
Meinshausen and Bühlmann introduced stability selection in 2010 \cite{meinshausen2010stability,meinshausen2009p}, establishing it as a widely used method for stable variable selection and for controlling false positives. Stability selection improves the performance of a variable selection algorithm by applying this algorithm multiple times to random subsamples of the data, each of size $\lfloor n/2 \rfloor$. The variables selected are those chosen most frequently across subsamples.
\newline
\indent The main attractive aspect of stability selection comes from its control of false positives through an upper bound on the expected number of falsely selected variables, termed V for the rest of the article. However, this control holds under specific assumptions about the selection procedure, requiring it to surpass random guessing and a robust exchangeability condition concerning the selection of noise variables.
\newline
\indent To address the issues of these assumptions, a variation called Complementary Pairs Stability Selection (CPSS) was proposed by Rajen et al. \cite{shah2013variable} in 2013. CPSS divides variables into two groups based on their probability of selection: one with a low probability and the other with a high probability. Subsamples as complementary pairs are drawn, and the selection procedure is applied to each. The selection probability for each variable is estimated, representing the proportion of fitted models where the associated coefficient estimate differs from zero. The final selected variables are those with a high selection probability, exceeding a user-defined cut-off.
Stability selection can be used for Type I error control through the upper bound V. However, in practice, stability selection is often used with the aim of stabilizing the obtained results in terms of selected variables, simply by choosing some (non-optimal) selection threshold.
We will introduce stability selection to the IPF-lasso  for variable selection with error control.

\section{Simulation}

\subsection{Simulation design} 
In this simulation study, we choose to explore three stability selection thresholds: manually chosen at 70\%, manually chosen at 80\%, and finally the formally optimal threshold, determined by the r-concave bound proposed by Rajen et al. \cite{shah2013variable}.
This bound is a function of the stability selection threshold. Thus, by selecting an appropriate upper bound for the expected  number of false positive selections, one can calculate the optimal threshold. All parameters that go into this calculation are selected as described in Rajen et al. \cite{shah2013variable}.

The value of B, corresponding to the number of complementary pairs of subsamples is set to 50. Stability selection with these three different thresholds will be applied both to the regular LASSO and to the IPF-LASSO. This will create six different procedures to compare.
The error control, by the number of false positives, and the power of these six models will be evaluated on simulated data. 
Given the multitude of potential simulation configurations, a lot of settings could be tested out. To make the comparison more comprehensible, we chose to design our simulation study very close to the one in the original IPF-LASSO article\cite{boulesteix2017ipf}.
\newline
\indent We simulate situations with two data modalities. Each modality is characterized by two parameters: the number of variables in each modality ($p_1$ and $p_2$) and the number of non-zero coefficients in each modality ($b_1$ and $b_2$). We chose a fixed signal strength of $\beta=1$ for both modalities across all simulations. Each design had a sample size of $n = 100$, and we generated 100 simulated datasets for each configuration. For all designs, we used a binary outcome. All considered designs in this simulation study are detailed in Table \ref{tab:Design}.

\begin{table}[t]
\begin{center}
\begin{tabular}{||c c c c c ||} 
 \hline
  & $p_1$ & $p_2$ & $b_1$ & $b_2$ \\ [0.5ex] 
 \hline\hline
 Design A & 1000 & 1000 & 10 & 10\\ 
 \hline
 Design B & 100 & 1000 & 3 & 30 \\
 \hline
 Design C & 100 & 1000 & 10 & 10\\
 \hline
 Design D & 100 & 1000 & 20 & 0 \\
 \hline
 Design E & 20 & 1000 & 3 & 10 \\ 
 \hline
 Design F & 20 & 1000 & 15 & 3 \\ 
 \hline
 Design G & 20 & 1000 & 10 & 10 \\ 
 \hline
 Design H & 20 & 1000 & 3 & 3 \\  
 \hline
 Design I & 20 & 2000 & 20 & 0 \\ 
 \hline
\end{tabular} 
\caption{\label{tab:Design} Configurations of the design used in the simulations}
\end{center}
\end{table}
These particular designs were considered to be more or less realistic, as in the situation of “useless omics data” where one of the modalities does not contain non-zero variables. 
\newline
\indent We considered two data settings: the first, called ``independent data", with no correlation within or between the two modalities, and the second, called ``correlated data", with a correlation both within and across modalities in the same dataset. This second setting with correlated data is obviously considered to be closer to real data. \\

The ``independent data" are simulated as follows:
\begin{enumerate}
\item The binary outcome is sampled from the Bernoulli distribution ${\mathcal {B}}(0.5)$.
\item The variables of the datasets are sampled from the multivariate normal distribution: \\
$X_1, X_2, \ldots, X_{p_1+p_2}| Y=0 \sim \mathcal{M}\mathcal{N} (0_{p_1+p_2} ,\Sigma)$\\
$X_1, X_2, \ldots, X_{p_1+p_2}| Y=1 \sim \mathcal{M}\mathcal{N} (\mu ,\Sigma)$\\
with $\mu^T=\left(\overbrace{ \underbrace{ \beta, \ldots,\beta}_{\text{length } b_1}, 0, \ldots, 0, }^{\text{length } p_1} \overbrace{\underbrace{ \beta, \ldots,\beta}_{\text{length } b_2}, 0, \ldots, 0}^{\text{length } p_2} \right)$\\
and $\Sigma=\mathrm {I} _{p_1+p_2}={\begin{pmatrix}1&0&\cdots &0\\0&\ddots &\ddots &\vdots \\\vdots &\ddots &\ddots &0\\0&\cdots &0&1\end{pmatrix}}$\\
\end{enumerate}

The ``correlated data" are simulated in the same way as the ``independent data" but the $\Sigma$ matrix is modified. To take into account both the correlation between variables of the same data modality and between variables of different modalities, the Identity matrix is replaced by a covariance matrix, assuming that there are b = 10 sets of variables that are mutually correlated within each modality. The correlation between variables belonging to the same block group is fixed at $\rho = 0.4$, and defined by the matrix $\Sigma$ as in Figure 1.

\begin{figure*}
\caption{Covariance matrix $\Sigma$ simulated for the ``correlated data" }
\setlength{\fboxsep}{0pt}%
\setlength{\fboxrule}{0pt}%
\begin{center}
\[
\Sigma = 
\left( \begin{array}{@{}c|c@{}}
   \begin{matrix}
      A_{p_1/b}(\rho) & & & {\fontsize{0.5cm}{0.5cm}\selectfont \text{0}} \\
      & A_{p_1/b}(\rho)& & \\
       & & \ddots & \\
  {\fontsize{0.5cm}{0.5cm}\selectfont \text{0}} & & & A_{p_1/b}(\rho)\\
   \end{matrix} 
      & \begin{matrix}
      B_{p_1/b,p_2/b}(\rho) & & {\fontsize{0.5cm}{0.5cm}\selectfont \text{0}}\\
      & \ddots&  \\
      {\fontsize{0.5cm}{0.5cm}\selectfont \text{0}} & & B_{p_1/b,p_2/b}(\rho) \\
  \end{matrix}  \\
   \cmidrule[0.4pt]{1-2}
   \begin{matrix}
      B_{p_2/b,p_1/b}(\rho) & & {\fontsize{0.5cm}{0.5cm}\selectfont \text{0}} \\
      & \ddots&  \\
      {\fontsize{0.5cm}{0.5cm}\selectfont \text{0}} & & B_{p_2/b,p_1/b}(\rho) \\
  \end{matrix} & \begin{matrix}
      A_{p_2/b}(\rho) & & & {\fontsize{0.5cm}{0.5cm}\selectfont \text{0}} \\
          & \ddots & \\
 {\fontsize{0.5cm}{0.5cm}\selectfont \text{0}}  & & A_{p_2/b}(\rho)\\
   \end{matrix}  \\
\end{array} \right)
\]
\end{center}
\end{figure*}

\begin{figure*}
    with $A_{q}(\rho)$ and $B_{q,q}(\rho)$ are $(q \times q)$ matrix defined as follow:
$A_{q}(\rho)=\begin{pmatrix}
1 &  & \rho \\
 & \ddots & \\
\rho &  & 1\\
\end{pmatrix}
$
and 
$B_{q,q}(\rho)=\begin{pmatrix}
\rho &  & \rho \\
 & \ddots & \\
\rho &  & \rho\\
\end{pmatrix}
$.
\end{figure*}

In order to randomly distribute the non-zero variables over the correlated blocks, the columns are randomly permuted after generating the data. These two simulation settings are the ones proposed in the original article about the IPF-LASSO.

We applied the six procedures presented previously to these simulation designs. For the IPF-LASSO method, the 11 following combinations of penalty factors were tested: 
\begin{center}
  $(1,1), (1,2), (1,4), (1,8), (1,16), (1,32), (1,64),$ $(1,\frac{1}{2}), (1,\frac{1}{4}), (1,\frac{1}{8}), (1,\frac{1}{16})$.   
\end{center}
The selection of candidate values for the tuning parameters is often arbitrary; however, the chosen combinations are based on the experience of Boulesteix et al. With the availability of bigger computational resources, the grid of combinations could be significantly refined. To choose the reference parameter $\lambda_1$ and the best combination of penalty factors, 5-fold cross-validation repeated 10 times is performed \cite{boulesteix2013complexity}, and because of the binary outcome, the error measure computed in the cross-validation is the classification error. No constraint is set on the maximal number of predictors allowed in the final model, i.e., the best model is selected based on CV without restriction. For the standard LASSO, the parameter $\lambda$ is determined using 5 folds cross validation repeated 10 times.
\newline
\indent In stability selection, the two first thresholds are chosen at 0.7 and 0.8 (70\% and 80\% selection probability). In the third stability selection configuration, we determine the threshold using the r-concave bound with V = 2 for the number of false positives, denoted as the ``optimal threshold" for the rest of the article. The value of this ``optimal threshold" can vary across simulation configurations but remains consistently between 0.5 and 0.55 in our case. This also means that the two first thresholds lead to very strict error control.
\newline
\indent To compare the performance of the six procedures, we compared the number of false positives in relation to the upper bound V = 2 . To evaluate the effectiveness of the selection procedures, we calculated the true positive proportion (TPP), also known as power. 

\subsection{Simulation results for ``independent data"}
\begin{figure*}[ht]
\caption{\label{fig:Fig2} Boxplot of the true positive proportion for all 9 designs and all 6 models for the independent data. The models are presented from left to right in the same order as in the legend (LASSO70: Lasso with threshold at 70\%, IPF\_LASSO80: IPF-Lasso with threshold at 70\%, LASSO80: Lasso with threshold at 80\%, IPF\_LASSO80: IPF-Lasso with threshold at 80\%, LASSO\_OPTI: Lasso with optimal threshold, IPF\_LASSO\_OPTI: IPF-Lasso with optimal threshold)}
\centering
\includegraphics[width=1.05\textwidth]{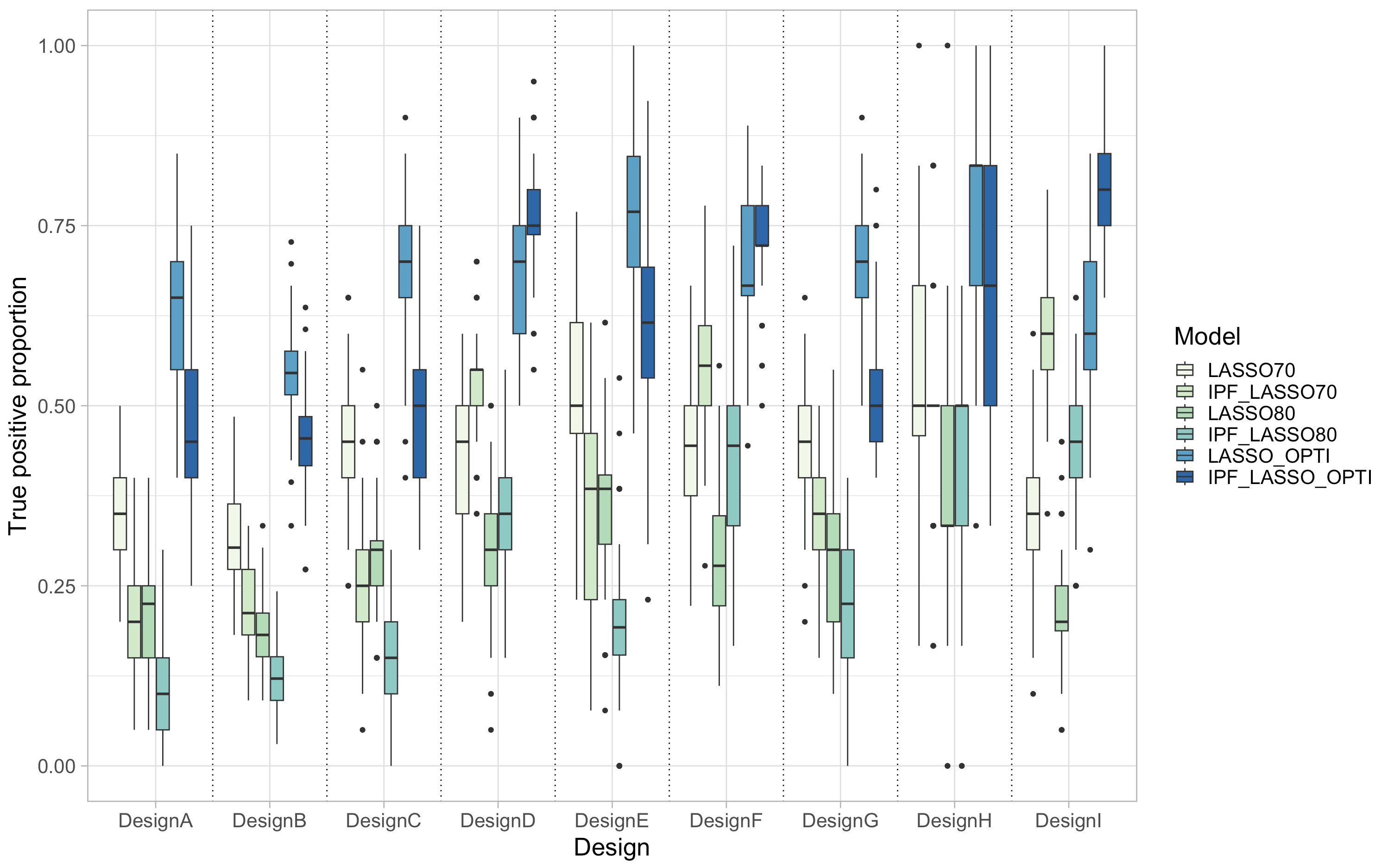}
\end{figure*}

\begin{figure*}[ht]
\caption{\label{fig:Fig3} Boxplot of the number of false positives for all 9 designs and all 6 models for the independent data. The models are presented from left to right in the same order as in the legend (LASSO70: Lasso with threshold at 70\%, IPF\_LASSO80: IPF-Lasso with threshold at 70\%, LASSO80: Lasso with threshold at 80\%, IPF\_LASSO80: IPF-Lasso with threshold at 80\%, LASSO\_OPTI: Lasso with optimal threshold, IPF\_LASSO\_OPTI: IPF-Lasso with optimal threshold)}
\centering
\includegraphics[width=1.05\textwidth]{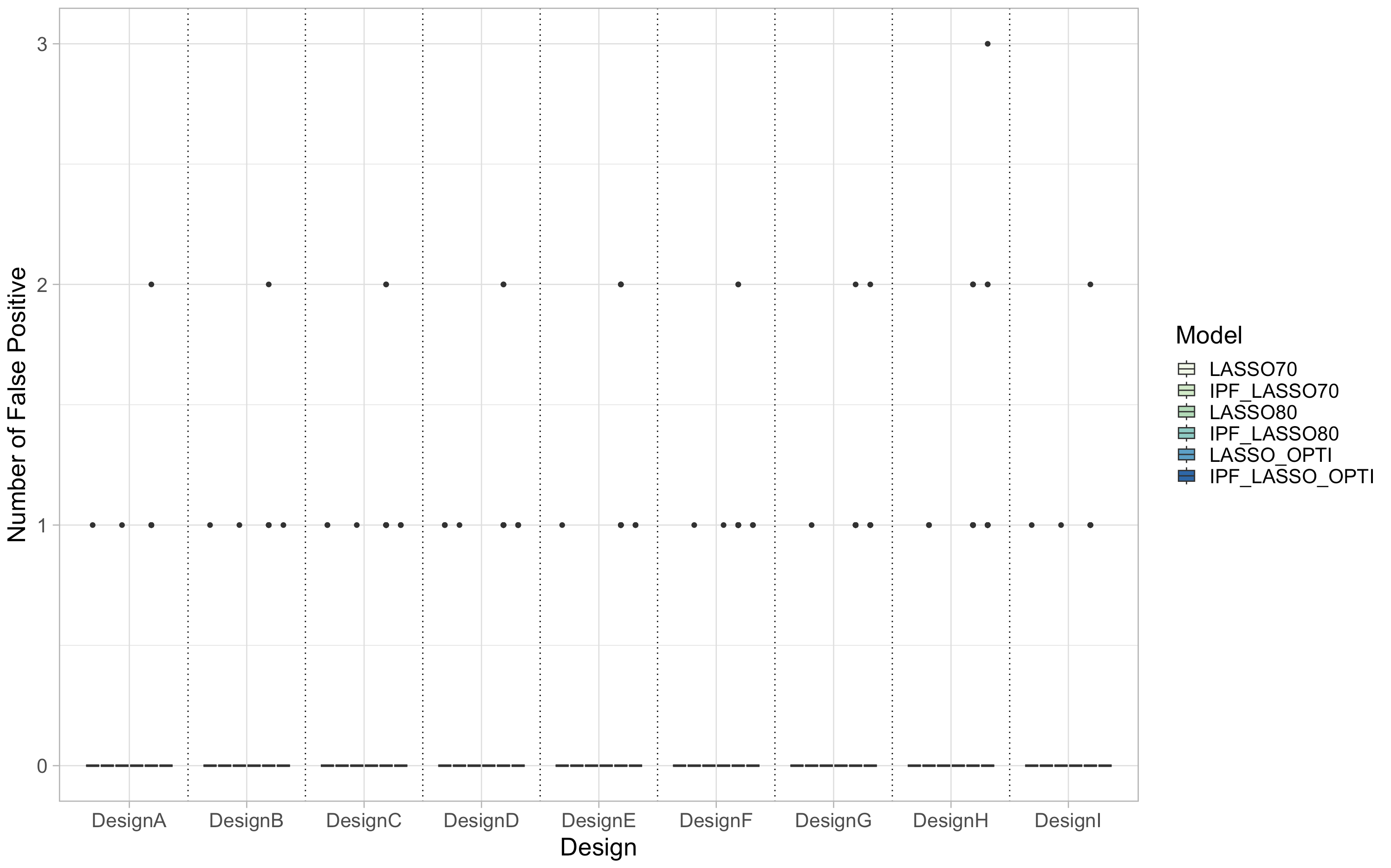}
\end{figure*}

For the ``independent data", the results of the nine simulated designs applied to the six procedures with the two criteria described in the previous section are shown respectively in Figure \ref{fig:Fig2} and in Figure \ref{fig:Fig3}. Figure \ref{fig:Fig2} and Figure \ref{fig:Fig3} represent the boxplot of all models. 

In all nine designs, all six procedures performed well in terms of controlling the number of false positives that we set to V = 2. Overall, very few false positives were selected at all in any of the experiments. This is no surprise, as the bound is known to be conservative. Since there is no difference in terms of false positives between the design and the procedures, it is interesting to look at the power (TPP). Here, the results are varying. In Design D, F and I, where the two datasets are very different in terms of the proportions of non-zero variables, the IPF-LASSO demonstrates superior power compared to regular LASSO. It is reasonable that when one of the modalities contains more informative variables, the different penalties introduced by the IPF-LASSO lead to higher power while maintaining error control. On the other hand, regular LASSO seems to outperform the IPF-LASSO in Designs A, B and C, where respectively, the two modalities, the proportion of truly relevant variables, and the number of truly relevant variables are the same. These results correspond to the results regarding prediction performance in the original IPF-LASSO paper.

\subsection{Simulation results for ``correlated data"}
\begin{figure*}[ht]
\caption{\label{fig:Fig4} Boxplot of the true positive proportion for all 9 designs and all 6 models for the correlated data. The models are presented from left to right in the same order as in the legend (LASSO70: Lasso with threshold at 70\%, IPF\_LASSO80: IPF-Lasso with threshold at 70\%, LASSO80: Lasso with threshold at 80\%, IPF\_LASSO80: IPF-Lasso with threshold at 80\%, LASSO\_OPTI: Lasso with optimal threshold, IPF\_LASSO\_OPTI: IPF-Lasso with optimal threshold)}
\centering
\includegraphics[width=1.05\textwidth]{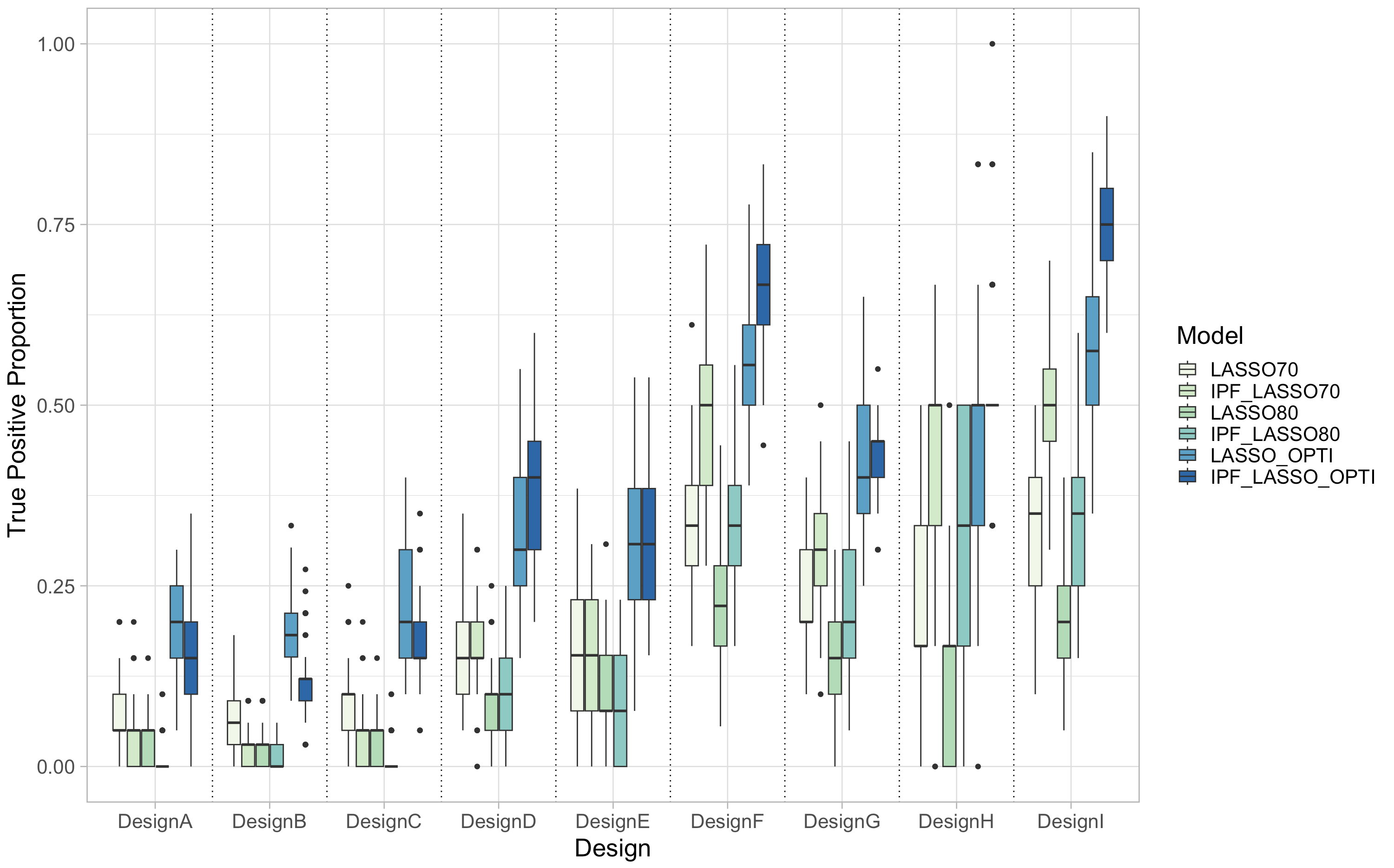}
\end{figure*}

\begin{figure*}[ht]
\caption{\label{fig:Fig5} Boxplot of the number of false positives for all 9 designs and all 6 models for the correlated data. The models are presented from left to right in the same order as in the legend (LASSO70: Lasso with threshold at 70\%, IPF\_LASSO80: IPF-Lasso with threshold at 70\%, LASSO80: Lasso with threshold at 80\%, IPF\_LASSO80: IPF-Lasso with threshold at 80\%, LASSO\_OPTI: Lasso with optimal threshold, IPF\_LASSO\_OPTI: IPF-Lasso with optimal threshold)}
\centering
\includegraphics[width=1.05\textwidth]{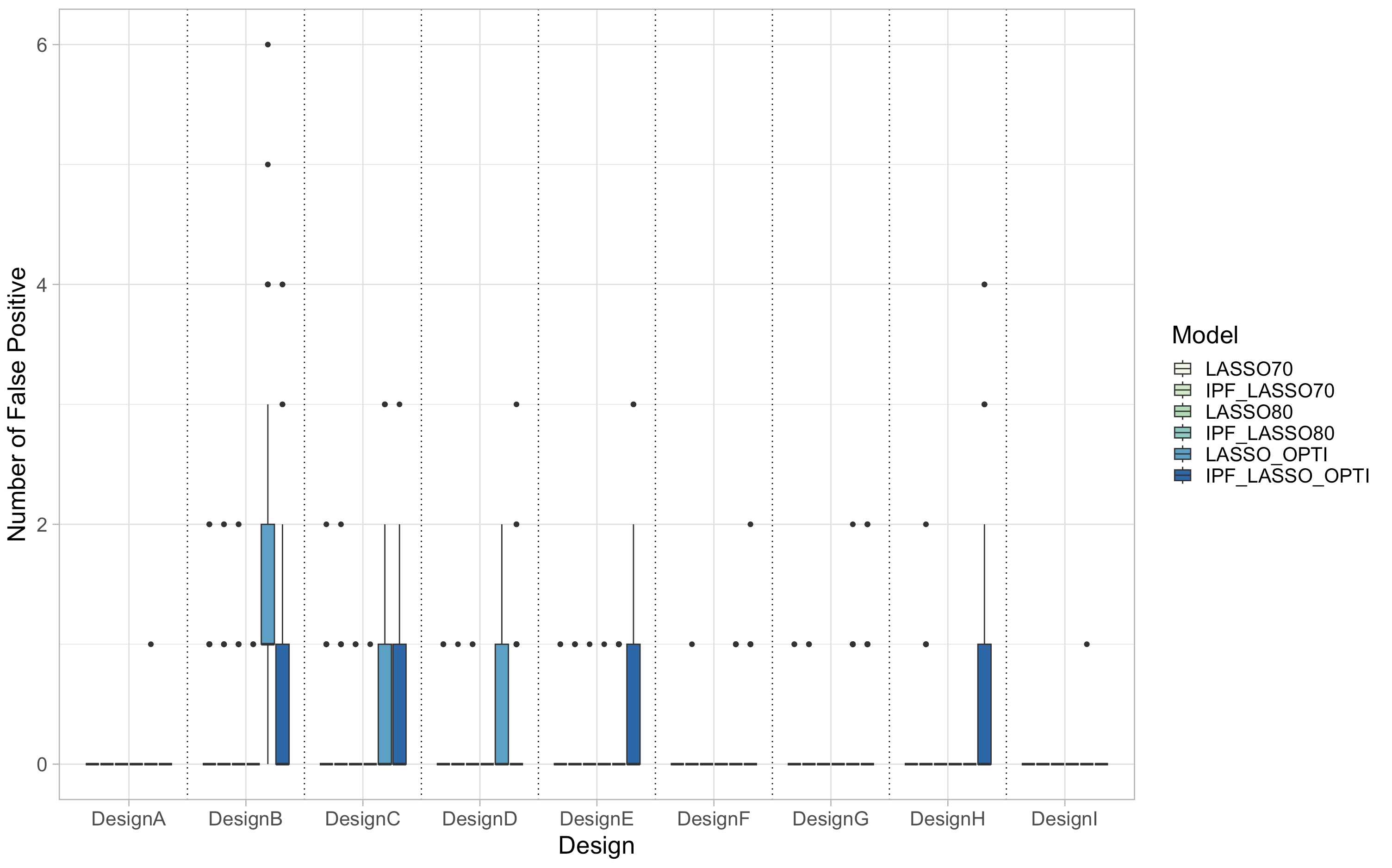}
\end{figure*}

For the ``correlated data”, the results of the nine simulated designs applied to the six procedures with the two criteria described in Section 3.1 are shown in Figure \ref{fig:Fig4} and Figure \ref{fig:Fig5}. Figure \ref{fig:Fig4} and Figure \ref{fig:Fig5} represent the boxplot of all models. 

The results for correlated data are very similar to the results obtained for independent data in terms of variable selection. These six procedures do not seem to be disrupted by the correlated data and they still perform well in terms of error control. The power is generally lower compared to the situation with independent data, as expected. Similar to the independent data scenario, IPF-LASSO outperforms regular LASSO in Designs D, F, and I, exhibiting higher power. In the correlated data context, IPF-LASSO also demonstrates superior power compared to regular LASSO in two additional designs: G and H. In Design H, there is little no variation in the results for the IPF-Lasso model with the optimal threshold. The boxplot is concentrated on 0.5, surrounded by some extremes values. In these two settings, the particular characteristics include a low number of true active variables (less than 0.5\%) in one of the two data modalities (in Design H) and a large size difference between the two modalities. It seems logical that when the size of the modalities differ significantly, the different penalties introduced by IPF-LASSO will balance the information given by both modalities. Giving them the same penalty could cause the larger modality to be favored at the expense of the smaller one, which could still contain relevant information.\

In summary, in this simulation study, we discovered three patterns in which the IPF-LASSO algorithm performs better than the regular LASSO. The first pattern is when the difference in proportions of non-zero variables between the two datasets is large. The bigger the difference, the better is the IPF-LASSO as compared to the regular LASSO. The second pattern which benefits from the IPF-LASSO is when the number of non-zero variables is small in all of the modalities. The last pattern in favour of the IPF-LASSO is the one with two modalities of very different sizes. The larger the difference between the the sizes, the better is the performance of the IPF-LASSO as compared to the regular LASSO. These three patterns are the same as those highlighted in the original article by Boulesteix et al. for prediction on multi-omics data. 

\section{Real Data Application}

\subsection{Breast Cancer Data}
Aligning with our idea to stay close to the work of Boulesteix et al. \cite{boulesteix2017ipf}, we use the same breast cancer data as in their article to illustrate our suggested variable selection procedure. The data has two modalities: one low-dimensional modality containing clinical data and one high-dimensional modality with gene expression data. We are interested in classification of patients with a minimal residual disease (RCB-I) vs. patients with a moderate or extensive residual disease (RCB-II and RCB-III) after chemotherapy. The data were initially analyzed by Hatzis et al. \cite{hatzis2011genomic}, and are publicly available through the Gene Expression Omnibus repository (GSE25066). We decided not to apply any preselection step, for two reasons: firstly, to make use all the information provided by all variables, and secondly, to avoid introducing bias by choosing among various preselection methods.
The clinical set contains the same variables as being used in the work of De Bin et al. \cite{de2014investigating}. 
The clinical variables are: 
\begin{itemize}
\item Age (continuous) 
\item Nodal status before treatment (4 categories): N0 (no lymph node involved), N1 (1-3), N2 (4–9), and N3 (10 and more).
\item Tumor size (4 categories): T1 (less than 2cm), T2 (between 2-5cm), T3 (more than 5cm), and T4 (any size in the chest) 
\item Grade (3 categories): Grade 1 (score 3,4,5),  Grade 2 (score 6,7), and Grade 3 (score 8,9) \cite{Breast}
\item Estrogen receptor (positive / negative)
\item Progesterone receptor (positive / negative)
\end{itemize}
For the analysis, the variables nodal status, tumor size and grade are represented by dummy variables in the model. The categories N0 (nodal status 0),  T1 (tumor size 1) and Grade 1 are used as reference categories. 

The high-dimensional modality contains 22,283 microarray gene expression measurements (Affymetrix-U133A GeneChip). More details about the data are available in the original paper \cite{hatzis2011genomic}.  After removing patients with missing data, the final dataset contains 385 patients, of which 277 are classified as RCB-I. 

\subsection{Analysis results}
We apply the six procedures presented previously with logistic regression on the breast cancer data. As for the simulated data, the following combinations of penalty factors were used for the IPF-LASSO:
\begin{center}
 $(1,1), (1,2), (1,4), (1,8), (1,16), (1,32), (1,64),$ $(1,\frac{1}{2}), (1,\frac{1}{4}), (1,\frac{1}{8}), (1,\frac{1}{16})$  
\end{center} 
and the best combination was chosen by cross-validation.
\newline
\indent Among the six procedures, only the IPF model with the optimal stability selection threshold (c=0.50) was able to select any variables. The upper bound was in this case set to V = 5. The only variable selected was age, which is not particularly interesting.  
However, it is well known
that with $\ell_1$-penalty, the presence of highly correlated variables can cause problems for stability selection \cite{kirk2013balancing}. Adding a small $\ell_2$-penalty, leading to the Elastic Net penalty\cite{zou2005regularization}, can alleviate this
issue. 
Thus, our six procedures were applied again to the real data with the Elastic Net penalty (the mixing parameter alpha set to 0.7). Again, only the IPF model with the optimal threshold for stability selection (c=0.5) was able to select any variables. Only clinical variables were selected. This is of course not very surprising, as the clinical variables are much ``closer" to the outcome (treatment response) compared to the gene expressions\cite{hatzis2011genomic}. The variables selected were Age, Estrogen Receptor, Nodal status, and Grade. 

\section{Conclusion}
This article focuses on an investigation of whether the IPF-LASSO method can achieve increased statistical power while controlling false positives compared to the standard LASSO approach in a multi-omics data settings. To our knowledge, this work is the first to treat variable selection with error control while integrating several different correlated data sources. Stability selection was implemented in both IPF-LASSO and regular LASSO on a binary outcome with the aim of controlling the number of false positives. 

Simulation studies highlight that, in scenarios where there are differences between the two data sources in terms of size of the data, quantity, or ratio of true variables, the IPF-LASSO method shows increased statistical power while controlling false positives compared to the conventional LASSO approach. In these scenarios, the IPF-LASSO should be preferred over the regular LASSO. 

On the cancer data, one clinical data modality and one omics data modality were used to predict the binary outcome and only the IPF-LASSO model was able to select any variables. All the selected variables were clinical variables. In this particular situation, this shows the superiority of the IPF-LASSO, which was capable of selecting variables even among the smallest dataset. The fact that no  gene expressions were selected could be explained by our choice of response variable (treatment response), which is obviously closely related to the clinical variables. However, we should be aware that our upper bound for false positives selection was higher than the number of variables that was actually selected.

While the IPF-LASSO appears advantageous for variable selection with false positive control in multi-omics data, like all LASSO-type methods, this method may not perform well when there are highly correlated predictors. An upgrade of the IPF-LASSO could be to substitute the LASSO penalty with an Elastic Net penalty, like we did in our applied example, see also Djordjilovic et al. \cite{djordjilović2024penalizedclr}. Comparing this modified ``IPF-elastic-net" model to the standard Elastic-net model for variable selection with false positive control in multi-omics data settings holds potential interest and merit for further exploration. 
This work is focused on controlling the number of false positives, but other types of error control exist, like the False Discovery Rate (FDR). Looking into ways of controlling FDR in the IPF-LASSO would be a natural extension of this work. Finally, continued progress in computational and statistical methods is crucial to achieve biologically meaningful integration of high-dimensional multi-omics data \cite{wissel2023systematic}.
\newline

\section*{Declaration of conflicting interests}
The author(s) declared no potential conflicts of interest with respect to the research, authorship, and/or publication of this article.

\section*{Funding}
The author(s) disclosed receipt of the following financial
support for the research, authorship, and/or publication of
this article: This research is funded by Norwegian Cancer
Society, Grant no. 216137.

\bibliographystyle{IEEEtran}
\bibliography{main}

\end{document}